\documentclass[conference]{IEEEtran}
\IEEEoverridecommandlockouts
\usepackage{cite}
\usepackage{amsmath,amssymb,amsfonts}
\usepackage{algorithmic}

\usepackage{graphicx}
\usepackage{textcomp}
\usepackage{xcolor}
\def\BibTeX{{\rm B\kern-.05em{\sc i\kern-.025em b}\kern-.08em
    T\kern-.1667em\lower.7ex\hbox{E}\kern-.125emX}}

\usepackage{graphicx}
\usepackage{caption}    % for \captionof
\usepackage{subcaption}
\usepackage{times}  % DO NOT CHANGE THIS
\usepackage{helvet}  % DO NOT CHANGE THIS
\usepackage{courier}  % DO NOT CHANGE THIS
\usepackage[hyphens]{url}  % DO NOT CHANGE THIS
\usepackage{graphicx} % DO NOT CHANGE THIS
\usepackage{caption} % DO NOT CHANGE THIS AND DO NOT ADD ANY OPTIONS TO IT
\usepackage{algorithm}
\usepackage{newfloat}
\usepackage{listings}
\DeclareCaptionStyle{ruled}{labelfont=normalfont,labelsep=colon,strut=off} % DO NOT CHANGE THIS
\floatstyle{ruled}
\newfloat{listing}{tb}{lst}{}
\floatname{listing}{Listing}

\usepackage{algorithm}
\usepackage{algorithmic}
\usepackage{amsmath}
\usepackage{multirow}
\usepackage{color}
\usepackage{booktabs}
\usepackage{array}
\usepackage{graphicx}
\usepackage{subcaption}
\usepackage{amsthm}
\theoremstyle{definition}
\newtheorem{example}{Example}
\usepackage{xcolor,colortbl}
\newcommand{\nop}[1]{}{}
 
\definecolor{Gray}{gray}{0.9}
\usepackage{todonotes}

\newcolumntype{a}{>{\columncolor{Gray}}c}
\NewDocumentCommand{\matt}
{ mO{} }{\textcolor{teal}{\textsuperscript{\textit{Matt}}\textsf{\textbf{\small[#1]}}}}

\begin{document}
\begin{sloppy}

\title{Learning to Attack: Uncovering Privacy Risks in Sequential Data Releases}

\author{\IEEEauthorblockN{Ziyao Cui\textsuperscript{*}}
\IEEEauthorblockA{\textit{Department of Computer Science} \\
\textit{Duke University}\\
Durham, NC, USA \\
richard.cui@duke.edu}
\and
\IEEEauthorblockN{Minxing Zhang\textsuperscript{*}}
\IEEEauthorblockA{\textit{Department of Computer Science} \\
\textit{Duke University}\\
Durham, NC, USA \\
minxing.zhang@duke.edu}
\and
\IEEEauthorblockN{Jian Pei}
\IEEEauthorblockA{\textit{Department of Computer Science} \\
\textit{Duke University}\\
Durham, NC, USA \\
j.pei@duke.edu}
}

\maketitle

\begin{abstract}
Privacy concerns have become increasingly critical in modern AI and data science applications, where sensitive information is collected, analyzed, and shared across diverse domains such as healthcare, finance, and mobility. While prior research has focused on protecting privacy in a single data release, many real-world systems operate under sequential or continuous data publishing, where the same or related data are released over time. Such sequential disclosures introduce new vulnerabilities, as temporal correlations across releases may enable adversaries to infer sensitive information that remains hidden in any individual release. In this paper, we investigate whether an attacker can compromise privacy in sequential data releases by exploiting dependencies between consecutive publications, even when each individual release satisfies standard privacy guarantees. To this end, we propose a novel attack model that captures these sequential dependencies by integrating a Hidden Markov Model with a reinforcement learning-based bi-directional inference mechanism. This enables the attacker to leverage both earlier and later observations in the sequence to infer private information. We instantiate our framework in the context of trajectory data, demonstrating how an adversary can recover sensitive locations from sequential mobility datasets. Extensive experiments on Geolife, Porto Taxi, and SynMob datasets show that our model consistently outperforms baseline approaches that treat each release independently. The results reveal a fundamental privacy risk inherent to sequential data publishing, where individually protected releases can collectively leak sensitive information when analyzed temporally. These findings underscore the need for new privacy-preserving frameworks that explicitly model temporal dependencies, such as time-aware differential privacy or sequential data obfuscation strategies. \footnote{Our implementation and experimental code are publicly available at \url{https://github.com/richardcui18/sequential-data-attack}.}
\end{abstract}


\begingroup
  \renewcommand\thefootnote{\fnsymbol{footnote}}% use symbol footnote
  \footnotetext[1]{Both authors contributed equally to this research.}
\endgroup

\section{Introduction}

Data privacy has emerged as a critical concern across many AI and data science applications, such as healthcare~\cite{abouelmehdi2018big, sajid2016data}, education~\cite{khalil2025towards, vie2022privacy}, finance~\cite{abbe2012privacy, byrd2020differentially}, and social media~\cite{beigi2018privacy, li2023adversary}. In response to these challenges, extensive research has focused on protecting data privacy in data releases, such as a patient's identity in a hospital database~\cite{chong2022bridging, hu2023privacy}, a training dataset for an AI model~\cite{cui2025membership, carlini2022membership, xie-etal-2024-recall, carlini2023extracting, yeom2018privacy}, and user browsing logs in personalized recommendation systems~\cite{ge2021privitem2vec, zhang2021membership}, using techniques such as differential privacy~\cite{abadi2016deep, friedman2010data, dwork2006differential, dwork2008differential} and federated learning~\cite{li2020review, mammen2021federated, zhang2021survey, wen2023survey}. 

However, most existing privacy-preserving mechanisms are designed for a single, static data release, rather than for scenarios involving repeated or continuous disclosures. In many real-world systems -- such as mobility tracking, healthcare analytics, financial reporting, and Internet of Things (IoT) monitoring -- data is generated and released sequentially or in real time. A growing body of evidence shows that even datasets protected by strong anonymization or privacy-preserving techniques can still leak sensitive information when multiple releases are analyzed jointly. For example, the 2018 Strava Heatmap incident\footnote{\url{https://www.theguardian.com/world/2018/jan/28/fitness-tracking-app-gives-away-location-of-secret-us-army-bases}} exposed the locations of secret U.S. military bases after aggregated fitness-tracking data revealed soldiers’ jogging routes. Likewise, the Netflix Prize dataset, initially anonymized for a machine learning competition, was later deanonymized by cross-referencing user ratings with publicly available IMDb reviews\footnote{\url{https://medium.com/@EmiLabsTech/data-privacy-the-netflix-prize-competition-84330d01cc34}}. These cases illustrate that privacy breaches frequently arise not from a single data disclosure, but from the composition effect -- the accumulation and interaction of multiple releases over time -- which fundamentally challenges the robustness of traditional privacy mechanisms in dynamic, real-world settings.

With this, ensuring privacy across a sequence of releases remains a major challenge. A natural question arises: \emph{Can privacy across multiple releases be guaranteed if each individual release is well protected?} More concretely, suppose that in each release, the probability that an attacker can correctly infer sensitive information -- such as an individual's exact location -- is bounded by a threshold $\lambda > 0$. Does this guarantee still hold when the attacker observes the entire sequence of published data? Unfortunately, the answer is no once an attacker possesses even limited background knowledge that links the releases temporally.

\begin{example}[Motivation]\label{ex:motivation-trajectory}
The growing availability of mobility data enables valuable applications in transportation, public health, and urban planning. During COVID-19, companies like Google released fine-grained movement data to monitor and mitigate disease spread\footnote{\url{https://www.google.com/covid19/mobility/}}. However, disclosing detailed trajectories, such as GPS coordinates, raises serious privacy risks. The FTC, for instance, has sued data brokers for selling such data, which can expose visits to sensitive locations\footnote{\url{https://www.wired.com/story/the-ftc-may-finally-protect-americans-from-data-brokers/}}. To protect privacy, locations within the trajectory are often coarsened to broader regions rather than exact coordinates.

\begin{figure}[t]
\centerline{\includegraphics[width=80mm]{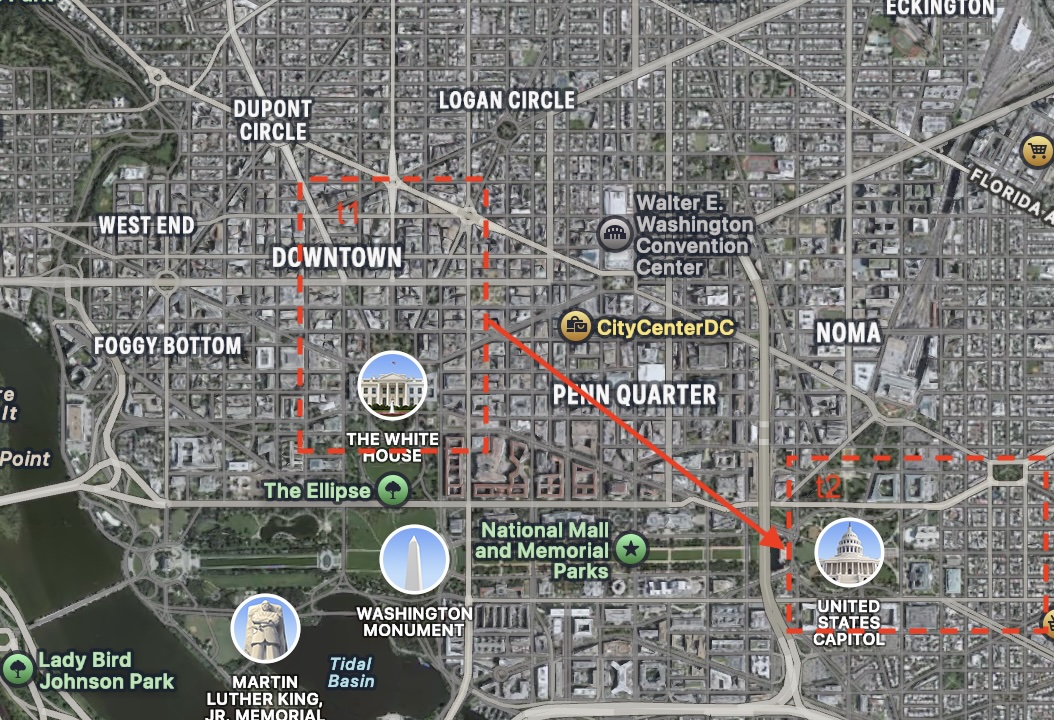}}
\caption{An illustration of sequentially published coarse-grained trajectories.\label{fig:motivation}}
\end{figure}

However, even sequences with every individual location protected with coarse location can inadvertently reveal sensitive information. Consider the example in Figure~\ref{fig:motivation}, where a user's trajectory at times $t_1$ and $t_2$ is published with each specific location becoming coarse-grained region (as marked in red) to preserve privacy. Looking at time $t_1$ individually, the likelihood that the user is near the White House may appear low since the region is large. However, considering $t_1$ and $t_2$ sequentially may cause privacy leakage: knowing that the user is in a region containing the Capitol at time $t_2$ substantially increases the probability that the user was sightseeing at the White House at $t_1$ and later visited the Capitol at $t_2$, given the background knowledge that tourists often visit those two locations in sequence. The user's privacy in the published sequence is therefore compromised.
\qed
\end{example}

The key idea illustrated in this example is that, with some background knowledge, an attacker can compromise privacy across multiple sequential data releases, even when each individual release appears safe. This observation motivates our study. Yet, it remains unclear how to model and quantify an attacker's background knowledge in such sequential settings, or how to automate an attack that exploits temporal correlations effectively.

At a high level, our approach builds on the intuition that human or system behaviors often follow sequential patterns that can be learned statistically. If each released dataset reveals a noisy or coarse view of the underlying truth, then the correlations between releases can help an adversary reconstruct the hidden sequence. We capture this intuition using a \emph{Hidden Markov Model (HMM)} to represent the latent true data (e.g., exact locations) and a \emph{reinforcement learning (RL)} framework to iteratively refine the attacker's model based on feedback from observed data. The HMM encodes the temporal dynamics -- the likelihood of transitioning between latent true data -- while the RL component adjusts the model parameters to improve inference accuracy over time. By combining these two elements in a \emph{bi-directional} framework, the attacker can reason not only forward (from past to future) but also backward (from future to past), effectively leveraging the full temporal context to uncover private information that individual releases alone would conceal.

In this paper, we focus on privacy attacks in sequential data publishing and investigate whether an adversary can compromise privacy by exploiting dependencies across multiple data releases. We make the following contributions:
\begin{enumerate}
    \item We formalize the problem of privacy attacks over sequential releases, showing that privacy guarantees preserved in individual releases may not hold when data is disclosed over time. 
    \item We propose a novel attack model that integrates a Hidden Markov Model with reinforcement learning to infer sensitive information from published sequences by leveraging both past and future contexts.
    \item We design a mechanism to represent and quantify the attacker’s background knowledge, enabling effective inference across temporally correlated data. 
    \item We demonstrate that our model substantially outperforms existing baselines through extensive experiments on both real and synthetic datasets, revealing a critical and underexplored privacy risk in sequential data publishing.
\end{enumerate} 

Beyond these technical contributions, our findings have broader implications for the design of privacy-preserving systems. They suggest that privacy guarantees must account for temporal dependencies and that privacy budgets or anonymization strategies should adapt dynamically as data evolves. Addressing these challenges requires rethinking privacy frameworks, auditing practices, and risk models for data that is continuously or periodically released.

The remainder of this paper is organized as follows. Section~\ref{sec: problem formulation} formulates the problem of privacy attacks against sequential data releases. Section~\ref{sec: related work} reviews related work on trajectory privacy preservation and privacy attacks. Section~\ref{sec: attack model} introduces our proposed attack model based on the Hidden Markov Model and reinforcement learning, while Section~\ref{sec: empirical result} presents experimental results and analysis on both real-world and synthetic datasets. Finally, Section~\ref{section: conclusion} concludes the paper and discusses potential directions for future research.
\section{Problem Formulation}
\label{sec: problem formulation}

In this section, we formulate the notion of privacy attacks against sequential releases. We start with the single time instant scenario and then extend to attacks over time.

\subsection{Single Time Instant Privacy Attacks}  

We begin with the simplest setting, where an attacker observes the published data at a single time instant. Consider a two-dimensional data space $\mathcal{D} = D_1 \times D_2$, where $D_1$ and $D_2$ are partitioned into intervals, and each interval serves as a basic unit of representation. With this, the data space can be viewed as a grid space consisting of multiple grid cells.

At a specific timestamp, the \textbf{true location} (assumed to be a single grid cell), denoted as $x$, of a trajectory can be expressed as a conjunctive normal form $V_{1} \wedge V_{2}$, where $V_j \in D_j$ for $j=1,2$ is a single interval. To protect privacy, the data publisher does not release the exact true location but instead discloses a \textbf{published region} that contains it. This published region can similarly be written as $U_{1} \wedge U_{2}$, where $V_{j} \in U_{j} \subseteq D_j$.  

Given the published region, an attacker may compute a \textbf{confidence score}, which is the probability that a specific grid cell corresponds to the true location. Since the true location is assumed to be a single grid cell (i.e., $|V_1| = |V_2| = 1$), the confidence given the true location $x$ is simply  
$
    \text{Conf}(x) = \prod_{j=1}^{2} \frac{1}{|U_j|}.
$  

To guarantee privacy, the data publisher ensures that this confidence does not exceed a user-specified threshold $\lambda > 0$, which is also known to the attacker. Formally, the publisher wants to ensure 
$
    \frac{1}{|U_1| \cdot |U_2|} \leq \lambda.
$  

\subsection{Privacy Attacks on One Trajectory Over Time}\label{sec:time_series_data_one_person}

In a privacy attack over time, an adversary observes a sequence of published regions rather than a single snapshot.  

Consider time instants $1, \ldots, T$ and define a \textbf{trajectory} 
$Y = \langle y_1, y_2, \ldots, y_T\rangle$, 
where $y_i = (t_i, x_i)$ is the $i$-th record of the trajectory, $t_i$ denotes the time instant, $x_i$, also denoted as $TL_i = V_{i,1} \wedge V_{i,2}$ in later discussions for better understanding, represents the object's \textbf{true location} at $t_i$, and $1 \leq t_1 < t_2 < \cdots < t_T \leq T$. For each time instant $t_i$, let $PR_i = U_{i,1} \wedge U_{i,2}$ 
denote the corresponding \textbf{published region}, where $TL_i \in PR_i$.  

The attacker's \textbf{confidence} in successfully attacking a true location $TL_i$ at any timestamp $t_i$ is defined as
$
   \text{Conf}(TL_i) =  \frac{1}{|U_{i,1}| \cdot |U_{i,2}|}.
$  
Thus, to ensure privacy, the data publisher requires that this confidence never exceed a pre-specified threshold $\lambda > 0$, which is assumed to be known to the public, including the attacker. Formally, for all $1 \leq i \leq T$, the publisher wants to ensure 
$
    \frac{1}{|U_{i,1}| \cdot |U_{i,2}|} \leq \lambda.
$  

From the attacker’s perspective, the goal is to reconstruct the sequence of true locations from the observed sequence of published regions. Formally, given $\langle PR_1, \ldots, PR_T\rangle$, the attacker seeks to learn a model  
\[
    \mathcal{F}(\langle PR_1, \ldots, PR_T\rangle ) = \widehat{TL} = \langle \widehat{TL}_1, \ldots, \widehat{TL}_T \rangle \in \mathcal{D}^T,
\]  
which outputs a predicted sequence of true locations $\widehat{TL}$ approaching the ground-truth sequence of true locations $TL=\langle TL_1,\ldots,TL_T\rangle$, where at time instant $t_i$, each $\widehat{TL}_i$ approaches its associated ground-truth true location $TL_i$.  

To quantify the prediction error, we employ the \textbf{average Euclidean distance (AED)} between the predicted and ground-truth sequences, that is,
$
    AED(TL, \widehat{TL}) = \frac{1}{T} \sum_{i=1}^T ED(TL_i, \widehat{TL}_i)
$, 
where $ED(\cdot,\cdot)$ denotes the Euclidean distance between two locations. Note that the Euclidean distance here can be replaced by other distance functions such as the Manhattan Distance flexibly.

For one trajectory, the attacker's objective is to minimize AED, subject to the constraint that the predicted true location at each time step lies within the corresponding published region:
\begin{equation}
    \begin{split}
       \min \quad & AED(TL, \widehat{TL})\\
        \text{s.t.} \quad & \widehat{TL}_i \in PR_i \quad \forall 1 \leq i \leq T 
    \end{split}
\end{equation}

\subsection{Privacy Attacks on Multiple Trajectories Over Time}\label{sec:time_series_data_person}

A data publisher may release trajectories for multiple objects over time. An important opportunity for an adversary is to exploit these multiple trajectories collectively, using them to mount coordinated attacks that compromise the privacy of several individuals simultaneously.  

Denote by $\mathcal{Y} = \{Y^{(1)}, \ldots, Y^{(S)}\}$ a set of $S$ trajectories, where each trajectory is represented as  
$
    Y^{(s)} = \langle (t_1^{(s)},TL^{(s)}_1), \ldots, (t_{T_s}^{(s)}, TL^{(s)}_{T_s}) \rangle,
$  
where $t_i^{(s)}$ denotes the $i$-th time stamp of trajectory $Y^{(s)}$ and $T_s$ denotes the length of trajectory $Y^{(s)}$ for $s = 1, \ldots, S$.  
The corresponding set of predicted trajectories is  
$
    \widehat{\mathcal{Y}} = \{\widehat{Y}^{(1)}, \ldots, \widehat{Y}^{(S)}\}
$,
where each $\widehat{Y}^{(s)} = \langle (t_1^{(s)},\widehat{TL}^{(s)}_1), \ldots, (t_{T_s}^{(s)},\widehat{TL}^{(s)}_{T_s}) \rangle$. A simple solution to obtain $\widehat{Y}^{(s)}$ is treating each trajectory individually and leverage the proposed model $\cal F$ illustrated in Section~\ref{sec:time_series_data_one_person}.

Given the ground-truth set of trajectories $\cal Y$ and predicted set of trajectories $\cal\hat{Y}$, we only extract the location information and thus denote $\cal L$ and $\cal\hat{L}$ as the set of sequences of ground-truth true locations and predicted true locations, respectively.

To evaluate prediction quality, we measure the error over all trajectories using the \textbf{aggregate average Euclidean distance (A$^2$ED)}:  
\[
    A^2ED(\mathcal{L}, \widehat{\mathcal{L}}) = \frac{1}{S} \sum_{s=1}^S AED\!\left(TL^{(s)}, \widehat{TL}^{(s)}\right).
\]
where $TL^{(s)}$ and $\widehat{TL}^{(s)}$ are the ground-truth and predicted sequence of true locations for trajectory $Y^{(s)}$, respectively.

The attacker's objective is to minimize this aggregate error, subject to the constraint that at every time step of each sequence, the predicted true location must lie within the corresponding published region:  
\begin{equation}
    \begin{split}
    \label{eq:traj utility}
        \min_{\widehat{\mathcal{L}}}\quad & A^2ED(\mathcal{L}, \widehat{\mathcal{L}}) \\
        \text{s.t.}\quad & \widehat{TL}_i^{(s)} \subseteq PR_i^{(s)}, \quad \forall\, 1 \leq i \leq T, \; 1 \leq s \leq S,
    \end{split}
\end{equation}
where $PR_i^{(s)}$ denotes the published region of trajectory $Y^{(s)}$ at time $t_i$.  

Alternatively, one can assess the \textbf{worst-case deviation} using the \textbf{aggregate maximum Euclidean distance (AMED)}:  
\[
    AMED(\mathcal{L}, \widehat{\mathcal{L}}) \;=\; \frac{1}{S} \sum_{s=1}^S \max_{i \in [1, T_s]} ED\!\left(TL^{(s)}_i, \widehat{TL}^{(s)}_i\right).
\]  
In this case, the optimization objective remains the same as in Equation~\ref{eq:traj utility}, except that $A^2ED$ is replaced with $AMED$. Both metrics are evaluated in our empirical study (Section~\ref{sec: empirical result}). Again, the Euclidean distance here can be replaced by other distance functions such as the Manhattan Distance flexibly.

\section{Related Work}
\label{sec: related work}

In this section, we briefly review prior research on trajectory privacy preservation methods and privacy attacks on anonymized trajectories.

\subsection{Trajectory Privacy Preservation Methods}
Trajectory privacy aims to protect location information over time. Existing approaches fall broadly into two categories.  

The first category is \textbf{synthetic trajectory generation}, which aims to produce realistic yet privacy-preserving mobility trajectories~\cite{hu2024real, yu2017seqgan, mohammed2024realistic}. Approaches in this category leverage a variety of generative models, including generative adversarial networks (GANs)~\cite{yu2017seqgan, feng2020learning, ouyang2018non, rao2020lstm, xu2021simulating, 10214943}, diffusion-based models~\cite{zhu2023difftraj, zhu2024controltraj, song2024controllable}, and large language models (LLMs)~\cite{zhang2025end, mohammed2024realistic}. These methods are designed to capture and reproduce the statistical characteristics of real-world mobility data -- such as the spatial distribution of visited locations, temporal transition patterns, and co-movement correlations -- while preventing the direct replication of individual trajectories.  

Despite these advances, recent studies~\cite{cecilia, 4497436, 9835666} reveal that synthetic data generation still faces substantial privacy risks due to \emph{memorization} effects during model training. In particular, Carlini et al.~\cite{carlini2023extracting} demonstrate that generative models can inadvertently memorize and reproduce sensitive training samples when trained with objectives aimed at aligning real and synthetic data distributions. Such memorization enables potential adversaries to extract private information from ostensibly anonymized outputs. These findings highlight a critical tension between realism and privacy in synthetic trajectory generation: models that are too faithful to real data risk compromising individual privacy, whereas overly obfuscated models lose their analytical utility. Consequently, ensuring rigorous privacy guarantees in synthetic trajectory generation remains an open and pressing research challenge.

The second category is \textbf{spatial generalization}, which focuses on modifying or obfuscating spatial information to protect individual privacy. One common direction is to release a real trajectory together with multiple synthetic or decoy trajectories, thereby concealing the true one among several plausible alternatives. Representative methods include $k$-anonymity, which ensures that each trajectory is indistinguishable from at least $k-1$ others~\cite{tu2018protecting, gramaglia2015hiding}. Another line of work perturbs true location data by introducing controlled randomness. Typical approaches employ \emph{differential privacy}, often implemented by injecting Laplace noise into spatial coordinates~\cite{jiang2013publishing, shao2020structured, zhu2024privacy}.  

However, because trajectory data are inherently high-dimensional and temporally correlated, achieving strong differential privacy guarantees often necessitates substantial noise addition, which significantly degrades data utility~\cite{jiang2013publishing}. To mitigate this trade-off, recent research has explored trajectory-specific adaptations of differential privacy~\cite{he2015dpt, jiang2013publishing, cao2019protecting, hua2015differentially, DBLP:journals/corr/abs-1112-2020, li2017achieving, liu2021differentially}. These tailored mechanisms aim to exploit the structural properties of mobility data to balance privacy and utility more effectively. Nevertheless, most existing methods assume independence between consecutive locations~\cite{6544872, 7930028}, an assumption rarely satisfied in real-world mobility datasets where strong temporal dependencies exist. This limitation restricts their effectiveness in capturing realistic movement patterns while maintaining rigorous privacy guarantees.

Recently, Zhang and Pei~\cite{10529533} proposed a greedy expansion method that hides true locations by publishing larger regions. While originally framed in the context of purchase-intent privacy in data market scenarios, this method also applies to trajectory data. However, their work primarily consider privacy within single, independent releases and overlook the cumulative risks introduced by temporal correlations. One of the key contributions of this paper is to attack this class of protection mechanisms. In short, to the best of our knowledge, no prior work has systematically investigated how sequential dependencies across multiple releases can undermine such region-based protection schemes. Our work is the first to formalize and evaluate privacy attacks that exploit these sequential dependencies, revealing that even when each individual release satisfies the intended privacy guarantees, sensitive information can still be inferred when the sequence is analyzed as a whole.

\subsection{Trajectory Attack Methods}
Attacks on anonymized trajectories can be grouped into \textbf{linkage} and \textbf{probabilistic} attacks~\cite{jin2022survey}.  

\textbf{Linkage attacks} re-identify individuals by combining anonymized trajectories with external data such as public transportation records and demographic information. With this combination, only knowing a few spatiotemporal points, an attacker can still re-link a trajectory to a specific individual~\cite{de2013unique}. Variants include record linkage (identity inference)~\cite{douriez2016anonymizing, riederer2016linking, maouche2017ap, jin2019moving}, attribute linkage (sensitive attributes inference)~\cite{gambs2010show, zang2011anonymization}, table linkage (membership inference)~\cite{pyrgelis2017knock}, and group linkage (social ties inference)~\cite{bilogrevic2013inferring, cho2011friendship}. These attacks, however, rely heavily on external auxiliary data or quasi-identifiers.  

\textbf{Probabilistic attacks} leverage confidence or uncertainty about hidden information for privacy attacks~\cite{gramaglia2017preserving, terrovitis2017local}. Under this type of attack, studies show that attackers can still recover sensitive trajectories even protected by differential privacy~\cite{jin2022survey}. For instance, reconstruction attacks exploit the structural distortions introduced by noise~\cite{buchholz2022reconstruction}, and sparsity-based approaches such as iTracker can recover multiple differentially private trajectories~\cite{8986753}.  

However, these prior attacks typically assume that each time instant is anonymized independently and that differential privacy is the primary privacy protection method, thereby overlooking both the sequential dependencies among locations and newly proposed region enlargement methods. In this work, we target the greedy expansion mechanism~\cite{10529533}, where individuals publish enlarged regions rather than exact true locations. We also explicitly exploit sequential dependencies in trajectories to infer true locations more accurately. 
% Earlier attack models~\cite{10529533} considered each time instant independently; we show that neglecting temporal correlations can lead to substantial privacy leakage, as discussed in Example~\ref{ex:motivation-trajectory}.

\section{Attack Model}
\label{sec: attack model}

In this section, we present our attack model that incorporates sequential information. Section~\ref{sec:attack strategy} highlights the intuition of the attack model. Section~\ref{subsec: baseline} introduces a baseline approach that performs attacks independently at each time instant and discusses its limitations. Section~\ref{subsec: hmm} then presents a Hidden Markov Model (HMM) as the fundamental framework for modeling sequential dependencies. Finally, Section~\ref{subsec:rl based approach} illustrates how reinforcement learning can be incorporated to further enhance the attack model.

\subsection{Attack Strategy}
\label{sec:attack strategy}

Given any trajectory $Y \in \cal{Y}$ (for simplicity, we omit the superscript $(s)$ from $Y$), since its ground-truth sequence of published regions $PR = \langle PR_1, \ldots, PR_T\rangle$ is publicly released and therefore available to an attacker, but the ground-truth sequence of true locations $TL$ is unavailable when training an attack model $\mathcal{F}(\langle PR_1, \ldots, PR_T\rangle)$, the attacker needs to rely on heuristics to assess prediction quality. One approach is to generate a predicted sequence of published regions $\widehat{PR}$ from the predicted sequence of true locations $\widehat{TL}$ and then compare $\widehat{PR}$ with the ground-truth sequence of published regions $PR$. Under this heuristic, the prediction quality of the attack model can be evaluated using the \textbf{Intersection over Union (IoU)} between the predicted and ground-truth sequence of published regions. Specifically, for each predicted $\widehat{PR}_i$ in $\widehat{PR}$ and associated ground-truth $PR_i$ in $PR$, the IoU can be computed as
$
    \text{IoU}(PR_i, \widehat{PR}_i) \;=\; 
    \frac{PR_i \cap \widehat{PR}_i}{PR_i \cup \widehat{PR}_i}
$. This predicted published region $\widehat{PR}_i$ is obtained from the predicted true location $\widehat{TL}_i$ together with a learned or assumed true-location-to-published-region (T2P) mapping available to the attacker. %Further details of this mapping will be discussed in Section~\ref{subsec:rl based approach}.

In practice, the adversary may obtain the T2P mapping through several strategies. A common approach is to assume a symmetric spatial expansion with respect to the true location until the privacy lower bound $\ell = \frac{1}{\lambda}$ is reached, followed by a small random displacement. This deterministic and weakly randomized policy is often referred to as greedy expansion~\cite{10529533}. Other approaches achieve a similar goal by publishing coarser regions given a trajectory to achieve $k$-anonymity~\cite{sweeney2002k, yu2023trajectory, chen2020differential}. Additionally, attackers may leverage expectation-maximization procedure~\cite{baum1970maximization} to learn a probabilistic T2P model, estimating the conditional distribution $P(\widehat{PR}_i|\widehat{TL}_i)$.

Note that the attacker’s assumed T2P mapping need not exactly coincide with the publisher’s actual mapping; it is used primarily to dramatically reduce the search space. As demonstrated in Section~\ref{sec: empirical result}, our experiments confirm that even when the attacker’s T2P differs from the real publishing strategy, a plausible T2P model can still substantially aid inference.

\subsection{Baseline Approach}
\label{subsec: baseline}

Given that the attacker can access the ground-truth sequence of published regions for each trajectory, a natural baseline strategy is to guess randomly at each time step, treating all time steps independently.  

Formally, for each time step $t_i$, the attacker observes the published region $PR_i$, which consists of a set of grid cells. The baseline strategy predicts the true location $\widehat{TL}_i$ by randomly picking one grid cell from the published regions $PR_i$. Under this strategy, the probability of correctly guessing the true location at time $t_i$ is simply $1/|PR_i|$, where $|PR_i|$ denotes the number of grid cells contained in the published region at $t_i$.  

This naive approach suffers from several limitations. First, the expected accuracy is typically very low, especially when the published region is large. More importantly, it ignores temporal dependencies in the trajectory. Intuitively, an object’s location at time $t_i$ is likely correlated with its locations at neighboring time steps $t_{i-1}$ and $t_{i+1}$. By treating each time step in isolation, the baseline approach fails to exploit this temporal structure. 
% thus provides only a weak benchmark for sequential privacy attacks.

\subsection{A Hidden Markov Model Approach}
\label{subsec: hmm}

Since the ground-truth true location of the object is unobserved while the ground-truth published region is observable, and given that each published region is derived from the corresponding true location at that time, we model the relationship between true locations and published regions using a Hidden Markov Model (HMM). In this formulation, the true locations are treated as the hidden states and the published regions as the observed states. The attacker's objective is to learn the transition and emission matrices of the HMM and to infer the most likely sequence of true locations given the observed (ground-truth) sequence of published regions.

At each time step, the hidden state corresponds to the object's true location. Because the true location is assumed to be a single grid cell within the published region, we define the hidden state space $\mathcal{H}$ as the union over all time steps of singleton subsets of the observed published regions:
\begin{equation*}
\begin{split}
    \mathcal{H} = \bigcup_{i=1}^T \left\{\widehat{V}_{i,1} \wedge \widehat{V}_{i,2}\;\middle|\;\widehat{V}_{i,j}\in U_{i,j}\ \forall j \in [1,2],\ \prod_{j=1}^2 |\widehat{V}_{i,j}|=1\right\},
\end{split}
\end{equation*}
where $U_{i,j}$ is the $j$-th dimension of the published region at time $i$, and $\widehat{V}_{i,j}$ is the $j$-th dimension of the object's predicted true location.

To define the observed state space, we first include all the ground-truth published regions $PR_1, \ldots, PR_T$. We then expand this set to account for plausible alternatives that the attacker might consider, given knowledge of the publisher’s privacy guarantees. Specifically, if the attacker knows the privacy threshold $\lambda$, the minimum published region size satisfying $\lambda$-level privacy is $\ell = \tfrac{1}{\lambda}$.

In principle, one could include all the sets that include the object's possible true locations whose sizes satisfy this constraint, but doing so would lead to prohibitive computational costs. Moreover, many large regions are unrealistic: a publisher aiming to preserve both privacy and utility is unlikely to release an overly and unnecessarily coarse region, since excessively large regions can severely reduce the usefulness of the data for downstream applications such as epidemiological modeling, transportation analysis, or urban planning. An arbitrary choice of published region size may therefore harm the balance between privacy protection and data utility.

To balance privacy preservation with practical utility, we introduce a size hyperparameter $\gamma$ that specifies a requirement on the usefulness of the anonymized data. In particular, $\gamma$ restricts the observed state space $\mathcal{O}$ to include only those published regions whose sizes are no more than $\gamma$ grid cells over the lower bound $\ell$. Formally,
\begin{equation*}
\begin{split}
    \mathcal{O} = \bigcup_{i=1}^T \left\{ \widehat{U}_{i} \;\middle|\; \widehat{U}_{i} \ni \widehat{V}_{i},\ \exists \widehat{V}_{i}\in \mathcal{H},\ \prod_{j=1}^2|\widehat{U}_{i,j}|\in [\ell, \ell+\gamma] \right\},
\end{split}
\end{equation*}
where $\widehat{V}_i = \widehat{V}_{i,1}\wedge \widehat{V}_{i,2}$ is one candidate true location from the attacker perspective, $\widehat{U}_{i,j}$ is the $j$-th dimension of a candidate published region (as we stated earlier, the observed state space not only include all the ground-truth published regions but also plausible alternatives) at time $i$, and $\widehat{U}_i = \widehat{U}_{i,1}\wedge \widehat{U}_{i,2}$ is the corresponding candidate published region.

\begin{figure*}[t!]
    \centering
    \includegraphics[width=1.0\textwidth]{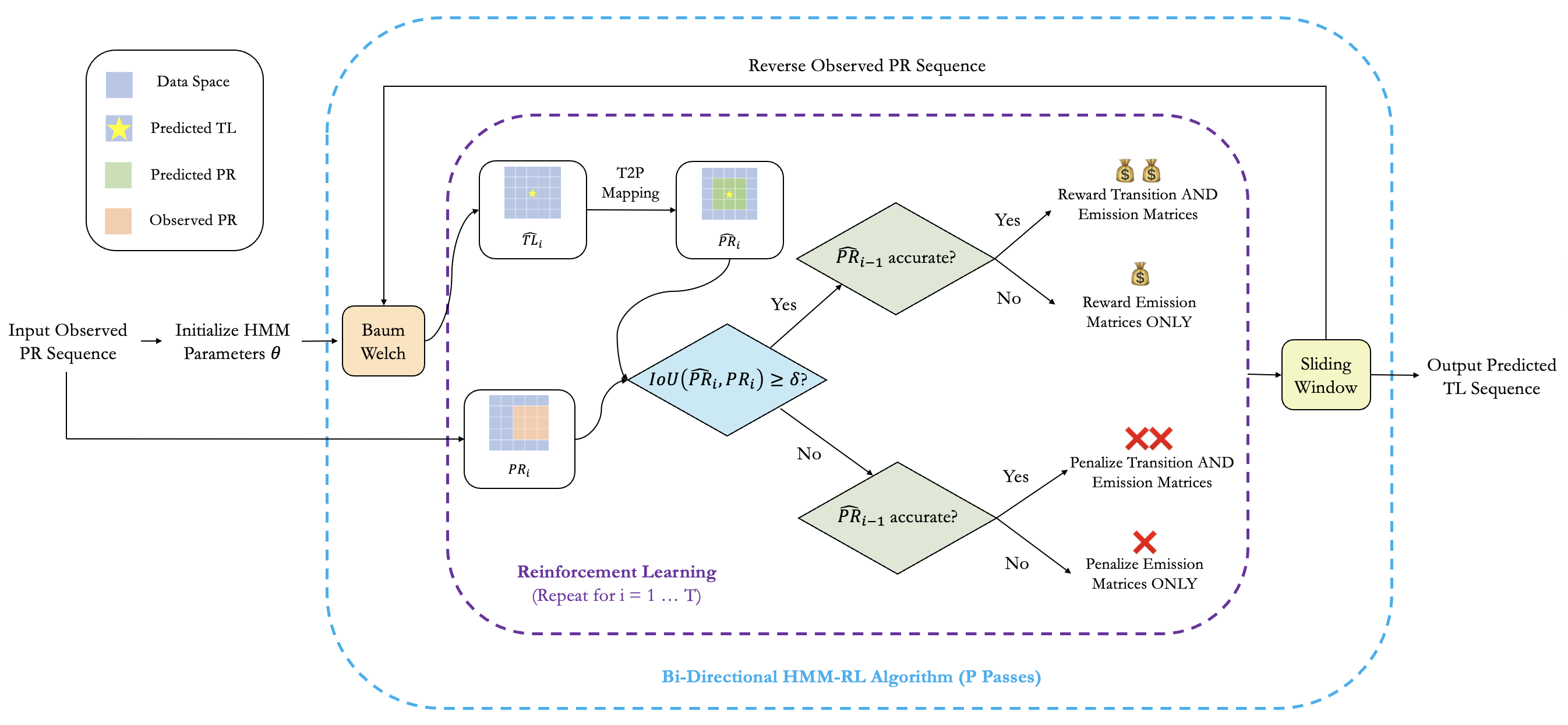} \\
    \caption{Overview of Bi-Directional HMM-RL Algorithm. $TL$ denotes the sequence of true locations, and $PR$ denotes the sequence of published regions.}
    \label{fig:hmm flowchart}
\end{figure*}

Given the constructed hidden and observed state spaces, we apply the Baum-Welch algorithm~\cite{baum1970maximization} to estimate the transition and emission matrices that maximize the likelihood of observing the ground-truth published region sequence. Once trained, we use the Viterbi algorithm~\cite{1450960} to infer the most likely sequence of hidden states, i.e., the predicted sequence of true locations. The predicted true location at any desired time step $i$ is then extracted from this sequence.

An important property of the above HMM training procedure is its ability to aggregate statistical evidence across all trajectories in the dataset. That is, instead of fitting each trajectory independently, the observed ground-truth sequences of all trajectories in $\mathcal{Y}$ are used to train the model, and thus the estimated transition and emission probabilities capture global mobility patterns that are shared among individuals, enabling the model to generalize beyond any single trajectory.

\subsection{A Reinforcement Learning-Based Bi-Directional Approach}
\label{subsec:rl based approach}

Beyond HMM, we also leverage reinforcement learning to improve the model performance. Since the ground-truth true locations are unavailable during training, we leverage the observed (ground-truth) sequences of published regions to guide reinforcement learning and refine the HMM-based model. In addition, we incorporate a bi-directional learning strategy that considers both past and future contexts, enabling the attack model to capture sequential dependencies more holistically. A schematic overview of the procedure is provided in Figure~\ref{fig:hmm flowchart} with the detailed illustrations as follows.

\subsubsection{Reinforcement Learning}

After running the Baum-Welch algorithm on the observed sequences of published regions, we can develop a heuristic of how the model performed. Let $\mathcal{F}$ denote the trained HMM and $\langle \widehat{TL}_1, \ldots, \widehat{TL}_T\rangle$ represent the predicted true locations over the time sequence $t \in [1, T]$ for one trajectory. Because the adversary does not have access to the ground-truth true location sequence $\langle TL_1, \ldots, TL_T\rangle$, direct evaluation of model accuracy is infeasible. However, the adversary can estimate the published region sequence $\langle \widehat{PR}_1, \ldots, \widehat{PR}_T\rangle$ from the predicted true location sequence using a T2P mapping as illustrated in Section~\ref{sec:attack strategy}, and then compare them against the observed (ground-truth) sequence of published regions $\langle PR_1, \ldots, PR_T\rangle$. To quantify the similarity between predicted and observed published regions, we employ the IoU metric $IoU(PR_i,\widehat{PR}_i)$ as defined in Section~\ref{sec:attack strategy}.

Since the attacker does not know how the data publisher generates their published region (i.e., the ground-truth T2P mapping), the attacker can only predict $\widehat{PR}_i$ using some learned or believed T2P mapping. A simple way for the attacker is to assume a T2P mapping in which the published region is centered on the predicted true location and satisfies the $\lambda$-level privacy constraint. In Section \ref{sec: sensitivity analysis}, we discuss the finding that, in general, attacks are more successful when the attacker's assumed T2P mapping used to obtain $\widehat{PR}_i$ is more similar to the ground-truth mapping used by the data publisher. However, even if the attacker assumed T2P mapping differs from the ground-truth mapping, experiments have shown that the attacks can still achieve high performance. 

% random guessing for each $i\in [1,T]$. Specifically, given a $\widehat{TL}_i$ state, the attacker would randomly choose one state that is a superset of $\widehat{TL}_i$ with size between the interval $[\ell, \ell+\gamma]$, where $\ell$ and $\gamma$ are defined in Equation \ref{eq:pi lower bound definition}. The randomly selected state would be used as $\widehat{PR}_i$ by the attacker.

Given the predicted published region $\widehat{PR}_i$, we use the IoU metric as the reward, denoted as $R_i$, to iteratively update the model parameters of $\mathcal{F}$ via reinforcement learning. High IoU values imply a strong alignment between predicted and observed published regions, indirectly suggesting that $\widehat{TL}_i$ is a plausible estimate of $TL_i$ (as $\widehat{PR}_i$ is directly inferred from $\widehat{TL}_i$ via the attacker assumed T2P mapping). However, the informativeness of $R_i$ is conditioned on the credibility of the previous true location estimate $\widehat{TL}_{i-1}$. That is, the rewards are only meaningful if $\widehat{TL}_{i-1}$ is accurate because if $\widehat{TL}_{i-1}$ is not accurate, then the transition probability can be correct even when the current $R_i$ is low, or incorrect even when the current $R_i$ is high. Accordingly, we define a threshold $\delta$ to assess the reliability of $R_{i-1}$, and update the transition probabilities as follows:
\begin{itemize}
    \item If $R_{i-1} \geq \delta$, then the predicted true location for $t_{i-1}$ is accurate, thus rewards are meaningful at time $t_i$. Thus, if $R_i \geq \delta$, we reward the transition probability $P(\widehat{TL}_i \mid \widehat{TL}_{i-1})$; if $R_i < \delta$, we penalize the transition.
    \item If $R_{i-1} < \delta$, then the predicted true location for $t_{i-1}$ is not accurate, thus we refrain from updating transition probabilities at time $i$.
\end{itemize}

\begin{algorithm}[t!]
\caption{Bi-Directional HMM-RL Algorithm\label{algo:main algo}}
\label{alg1}%\small
\begin{algorithmic}[1]
    \REQUIRE Set of $S$ sequences of observed (ground-truth) published region $\langle PR^{(1)}, \ldots, PR^{(S)} \rangle$, where $T_s$ is the length of sequence $PR^{(s)}$; T2P mapping.
    \REQUIRE Number of passes $P$, sliding-window size $k$, accuracy threshold $\delta$, privacy lower bound $\ell = \tfrac{1}{\lambda}$  
    \STATE Initialize model $\cal F$'s parameters $\theta \gets [$forward\_transition\_matrix, 
    backward\_transition\_matrix,  emission\_probability\_matrix, initial\_state\_distribution$]$ 
    \FOR{$pass \gets 1$ to $P$}  
        \IF{$pass$ is odd}
            \STATE Run Baum-Welch on $\langle PR^{(1)}, \ldots, PR^{(S)} \rangle$ with prior $\theta$, updating forward\_transition\_matrix, emission\_probability\_matrix, and initial\_state\_distribution only.
        \ELSE
            \STATE Run Baum-Welch on $\langle PR^{(S)}, \ldots, PR^{(1)} \rangle$ with prior $\theta$, updating backward\_transition\_matrix, emission\_probability\_matrix, and initial\_state\_distribution only.
        \ENDIF
        \STATE \quad \textit{Output: $\cal F$'s updated parameters $\theta$}  
        \FOR{$s \gets 1$ to $S$}  
            \STATE $\widehat{TL}^{(s)}$ $\gets$ $\cal F$.predict($PR^{(s)}$)  %\Comment{Predicted sequence of true locations.}
            % \STATE prev\_day\_accurate $\gets$ False  
            \FOR{$i \gets 1$ to $T_s$}  
                \STATE $\widehat{TL}_{i}^{(s)} \gets \widehat{TL}^{(s)}[i]$  
                \STATE $\widehat{PR}_{i}^{(s)} \gets T2P(\widehat{TL}_{i}^{(s)}, \ell)$ 
                % \COMMENT{Emission probability matrix can also be replaced with a T2P mapping if available.}  
                \STATE $R_i^{(s)} \gets \text{IoU}(\widehat{PR}_{i}^{(s)}, PR_{i}^{(s)})$  
                \IF{$R_{i-1}^{(s)} \geq \delta$ \AND $R_i^{(s)} \geq \delta$}  
                    \STATE Reward forward\_transition\_probability\_matrix or backward\_transition\_probability\_matrix.
        %             \IF{$pass$ is odd}
        %     \STATE Reward forward\_transition\_probability\_matrix.
        % \ELSE
        %     \STATE Reward backward\_transition\_probability\_matrix.
        % \ENDIF
        %$P(\widehat{TL}_i \mid \widehat{TL}_{i-1})$  
                \ELSIF{$R_{i-1}^{(s)} \geq \delta$ \AND $R_i^{(s)} < \delta$}  
                    \STATE Penalize forward\_transition\_probability\_matrix or backward\_transition\_probability\_matrix.%$P(\widehat{TL}_i \mid \widehat{TL}_{i-1})$  
                \ENDIF  
                \IF{$R_i^{(s)} \geq \delta$}  
                    \STATE Reward emission\_probability\_matrix 
                \ELSE
                    \STATE Penalize emission\_probability\_matrix %$P(\widehat{TL}_i \mid \widehat{TL}_{i-1})$  
                \ENDIF  
                % \STATE Reinforce emission\_probability\_matrix based on $R_i$  
                % \STATE prev\_day\_accuracy $\gets (R_i^{(s)} \geq \delta)$  
            \ENDFOR  
        \ENDFOR  
        \IF{$pass$ is odd}
            \STATE backward\_transition\_matrix $\gets$ average of backward\_transition\_matrices from last $k$ passes.
        \ELSE
            \STATE forward\_transition\_matrix $\gets$ average of forward\_transition\_matrices from last $k$ passes.
        \ENDIF
        % \STATE Apply sliding-window averaging over the last $k$ transition matrices in the same direction  
        % \STATE Reverse the observed region sequence  
        % \STATE Swap forward and backward transition matrices in $\theta$  
    \ENDFOR  
    \RETURN $\cal{F}$.predict($\langle PR^{(1)}, \ldots, PR^{(S)} \rangle$)
\end{algorithmic}
\end{algorithm}  

For emission probabilities, it shares the same logic when $R_{i-1}\geq \delta$ because the rewards are meaningful when the predicted true location for $t_{i-1}$ is accurate. When $R_{i-1} < \delta$, however, we believe that reinforcement learning is still meaningful. Assuming that $R_{i-1} < \delta$, consider the two possible cases: (1) $\widehat{TL}_i = TL_i$ and (2) $\widehat{TL}_i \neq TL_i$. For the first case, the emission probability $P(\widehat{PR}_i \mid \widehat{TL}_i)$ should still be rewarded or penalized according to $R_i$. For the second case, we could either reinforce $P(\widehat{PR}_i \mid \widehat{TL}_i)$ according to $R_i$ or do nothing. Although reinforcing $P(\widehat{PR}_i \mid \widehat{TL}_i)$ in this case might lead to potentially inaccurate adjustment of emission probabilities, we choose to still reinforce $P(\widehat{PR}_i \mid \widehat{TL}_i)$ for the following reason: not reinforcing the emission probabilities may theoretically result in a lack of parameter updates across iterations. For example, if all of $R_1, \ldots, R_T$ are less than $\delta$ during the first iteration, then without reinforcement on the emission probabilities, the same model parameters would be used in the second iteration, yielding identical predictions since no updates have occurred. As shown empirically through our experiments in Section~\ref{sec:effectiveness comparison}, disabling reinforcing emission probability when $R_{i-1} < \delta$ leads to noticeably worse model performance.

\subsubsection{A Bi-Directional Approach}

To further improve model performance, we adopt a bi-directional learning strategy that considers both forward and backward sequences of the observed published regions. That is, in addition to modeling the sequence $\langle PR_1, \ldots, PR_T\rangle$, we also leverage the reversed sequence $\langle PR_T, \ldots, PR_1\rangle$.  

A standard Hidden Markov Model (HMM) trained on a forward sequence captures temporal dependencies in a \emph{unidirectional} manner by learning transition probabilities $P(\widehat{TL}_i \mid \widehat{TL}_{i-1})$ for all $i \in [2, T]$. Modeling forward dependencies is well-motivated in many mobility applications. For example, knowing that an individual is on a highway at time $t_{i-1}$ effectively constrains the feasible locations at time $t_i$ to those topologically connected to the highway network -- such as highway segments, exits, or interchanges -- since transitions can only occur at designated points.  

However, modeling only how the current location depends on past locations, as in a standard HMM, is often insufficient. This approach neglects the \emph{inverse dependencies} $P(\widehat{TL}_{i-1} \mid \widehat{TL}_i)$, which can be equally informative in structured spatiotemporal systems. In many real-world scenarios, an entity’s current location is not determined solely by its past trajectory but can also be influenced by its anticipated or planned future states. For instance, an individual planning to attend a conference the following day may choose accommodation near the venue, making their current location a result of future intentions rather than preceding movements. Ignoring such bi-directional dependencies limits the model's ability to capture the full range of causal and anticipatory patterns inherent in human mobility behavior.

To capture such dependencies, we train a separate transition probability matrix on the reversed sequence of published regions, enabling the model to account for backward relationships. This reverse model complements the forward model and provides a more context-aware understanding of the underlying movement dynamics.  

\subsubsection{Final Model}
\label{subsec:final model}

We now present our complete \textbf{Bi-Directional Reinforcement Learning Hidden Markov Model} for sequential privacy attacks. The pseudo-code is given in Algorithm~\ref{algo:main algo}. 

The model maintains two transition matrices: a forward transition matrix $P(\widehat{TL}_i \mid \widehat{TL}_{i-1})$ capturing dependencies from $t_{i-1}$ to $t_i$, and a backward transition matrix $P(\widehat{TL}_{i-1} \mid \widehat{TL}_i)$ capturing dependencies from $t_i$ to $t_{i-1}$. Both directions share a common emission probability matrix.  

Training proceeds for $P$ passes, where $P$ is a tunable hyperparameter. Each pass updates transition and emission matrices using the Baum-Welch algorithm combined with reinforcement learning. Odd-numbered passes operate on the forward sequence $\langle PR_1, \ldots, PR_T\rangle$ and update only the forward transition matrix, while even-numbered passes operate on the reversed sequence $\langle PR_T, \ldots, PR_1\rangle$ and update only the backward transition matrix. To improve stability and convergence, we apply a sliding-window averaging scheme: at iteration $j$, the current transition matrix is initialized as the average of the most recent $k$ matrices from the same direction as illustrated in lines 26 to 30 in Algorithm~\ref{alg1}. For instance, if the current pass is odd, indicating the next pass will use backward transition matrix, we average the last $k$ backward transition matrices, where $k$ is another tunable hyperaparameter. If fewer than $k$ are available, no averaging is performed.  

This bi-directional reinforcement learning framework enables the attacker to exploit both past and future information when inferring true locations, yielding a more robust and context-aware model of sequential behaviors. 

\section{Empirical Results}
\label{sec: empirical result}

In this section, we evaluate the performance of the proposed Bi-Directional HMM-RL algorithm on both real and synthetic datasets. Section~\ref{sec:experimental setup} describes the experimental setup. Section~\ref{sec:effectiveness comparison} reports the overall effectiveness of the proposed method.
%, followed by a visual comparison in Section~\ref{sec: visualization results}
Finally, Section~\ref{sec: sensitivity analysis} explores the sensitivity of the model to key hyperparameters. %Section~\ref{sec: convergence analysis} further investigates algorithm convergence.

\subsection{Experimental Setup}
\label{sec:experimental setup}

\begin{table*}[t]
\centering
\caption{Overall effectiveness of the Bi-Directional HMM-RL algorithm on all datasets, measured using aggregate average Euclidean distance (A$^2$ED) and aggregate maximum Euclidean distance (AMED). \textbf{Bolded} entries denote the smallest Euclidean error across all models. EPRL denotes Emission Probability Reinforcement Learning applied when $R_{i-1} < \delta$.} %\matt{the full name of A^2ED and AMED}}
\label{table:overall results}
\begin{tabular}{lcccccccc} 
\toprule
& \multicolumn{2}{c}{Syn-Chengdu} & \multicolumn{2}{c}{Syn-Xi'an} & \multicolumn{2}{c}{Geolife} & \multicolumn{2}{c}{Porto Taxi} \\
\cmidrule(lr){2-3} \cmidrule(lr){4-5} \cmidrule(lr){6-7} \cmidrule(lr){8-9}
Model & A$^2$ED & AMED & A$^2$ED & AMED & A$^2$ED & AMED & A$^2$ED & AMED \\
\midrule
Baseline & 284.674 & 396.827 & 277.632 & \textbf{383.628} & 264.563 & 532.337 & 461.148 & \textbf{745.935} \\
HMM-RL (without EPRL) & 313.655 & 561.646 & 313.932 & 577.345 & 321.796 & 583.472 & 449.701 & 765.535 \\
\rowcolor{gray!20}
HMM-RL (with EPRL) & \textbf{195.987} & \textbf{388.490} & \textbf{197.546} & 389.179 & \textbf{204.068} & \textbf{427.527} & \textbf{360.259} & 749.464 \\
\bottomrule
\end{tabular}
\end{table*}

\subsubsection{Geolife Dataset} 
We first evaluate our method on the Geolife dataset~\cite{zheng2011geolife}, a two-dimensional, real-world trajectory dataset collected by Microsoft Research. The dataset contains 17,621 trajectories from 182 users between April 2007 and August 2012 all over the world. Each trajectory records GPS coordinates (longitude and latitude). The original trajectories are sampled every 1–5 seconds; we subsample every 18 seconds in our experiments, as time intervals that are too short often introduce noise or redundant stationary points, while time intervals that are too long tend to oversmooth the trajectories and obscure fine-grained movements.

Moreover, we choose a dense part of the data set for our experiments, which is the data collected in Beijing, China. We define a rectangular bounding box with longitude range [116.28, 116.32] and latitude range [39.95, 40.0], approximately 15 kilometers northwest of downtown Beijing. This area is discretized into grids of side length $\sim$99.383 meters. From this, we obtain 658 trajectories with lengths between 5 and 30 time steps.  

\subsubsection{Porto Taxi Dataset} 

We further evaluate our method on the Taxi Porto dataset\footnote{\url{https://www.kaggle.com/datasets/crailtap/taxi-trajectory/data}}, a large-scale real-world trajectory dataset that records one year of trips from all 442 taxis operating in the city of Porto, Portugal, between July 2013 and June 2014. Each trip is represented as a sequence of GPS coordinates sampled every 15 seconds, accompanied by contextual information including timestamps, day types, and indicators for missing data.

For our experiments, we define a rectangular bounding box covering the urban core of Porto and discretize it into uniform grids with a side length of approximately 148.957 meters. There are 1,710,671 trajectories in the dataset.

\subsubsection{Synthetic Datasets}
In addition to real datasets, we evaluate our method on SynMob~\cite{zhu2023synmob}, a high-fidelity synthetic GPS trajectory dataset. SynMob is generated using a diffusion-based trajectory synthesizer trained on large-scale proprietary mobility data, designed to preserve the key statistical and spatial-distributional properties of the original datasets while enabling rigorous analysis without access restrictions.

For our experiments, we focus on the Syn-Chengdu and Syn-Xi'an datasets. Each dataset contains one million synthetic trajectories represented in latitude-longitude format. In the Syn-Chengdu dataset, points are sampled every 3 seconds, covering the latitude range [30.65, 30.73] and longitude range [104.04, 104.13], which corresponds to a central area of Chengdu, China. This bounding box is discretized into grids with a side length of approximately 97.367 meters.

The Syn-Xi'an dataset is also sampled every 3 seconds, spanning the latitude range [34.20, 34.28] and longitude range [108.90, 108.99], corresponding to Xi'an, China. Its bounding box is discretized into grids with a side length of approximately 97.479 meters.

\subsubsection{Published Region Generation}
Since these datasets contain only exact locations and do not include the enlarged published regions used for privacy protection, we need to generate the published regions ourselves. For each sequence of true locations, we generate a sequence of published regions that satisfies the $\lambda$-privacy constraint. Specifically, for each true location, we first compute the minimum published region size $\ell = 1/\lambda$. Each published region is then initialized as a single grid cell centered on the true location. At each iteration, we \emph{randomly} select one axis (longitude or latitude) and expand the region symmetrically by one grid cell in both directions (east-west if longitude, north-south if latitude). This expansion continues until the published region reaches size $\ell$, ensuring the true location remains centered.  

To reflect realistic variability, we introduce a deviation parameter $d$, which shifts the published regions $d$ grid cells in a randomly chosen direction (east, west, north, or south) away from the true locations, ensuring that the true location does not always lie at the center of the published region. The deviated regions are then used as the observable published regions. Section~\ref{sec: sensitivity analysis} presents our model performance under varying values of $\lambda$ and $d$.  

\subsubsection{Implementation Details}
% \matt{move the later hyperparameter choices here and specify what is the baseline in the table}

We set $\lambda = 0.1$ and the deviation parameter $d = 2$ when generating the published region sequences. For the attack model, we adopt the following hyperparameters: $\delta = 0.7$, $\gamma = 5$, $k = 3$, and $P=50$. We compare the proposed model with a baseline model that attacks the true location at each time step independently. The detailed design of the baseline approach can be found in Section~\ref{subsec: baseline}.

\subsection{Effectiveness Comparison}
\label{sec:effectiveness comparison}

\subsubsection{Comparison With Baseline}
Table~\ref{table:overall results} presents the experimental results across all datasets. First, we compare the proposed Bi-Directional HMM-RL algorithm with EPRL against the baseline. Overall, HMM-RL with EPRL outperforms the baseline across both metrics, A$^2$ED and AMED, achieving average decreases of 22.28\% and 4.97\%, respectively. 

For A$^2$ED, HMM-RL with EPRL consistently outperforms the baseline across all datasets. For AMED, the baseline slightly outperforms HMM-RL with EPRL in two instances, but by only 4.54 meters on average. In contrast, when HMM-RL with EPRL is superior, it surpasses the baseline by an average of 73.809 meters.

The differences between the predictions produced by the Bi-Directional HMM-RL algorithm (with EPRL) and the baseline model become clearer when examining the geographic reconstruction of an example trajectory from the Geolife dataset, as shown in Figure~\ref{fig:one traj}. The ground-truth trajectory is depicted as a solid purple line, while the predicted trajectories are shown as dashed lines. Even when the deviation parameter is set to $d = 2$ during published region construction to hide true locations (corresponding to approximately 198.766 meters of additional obfuscation), our proposed HMM-RL with EPRL (in blue) still successfully reconstructs the majority of true locations and thus closely mimics the ground-truth trajectory. In contrast, the baseline predictions (in orange) deviate significantly from the ground-truth true locations.

\begin{figure*}[t!]
  \centering
  \begin{minipage}[b]{0.32\textwidth}
    \centering
    \includegraphics[width=\linewidth]{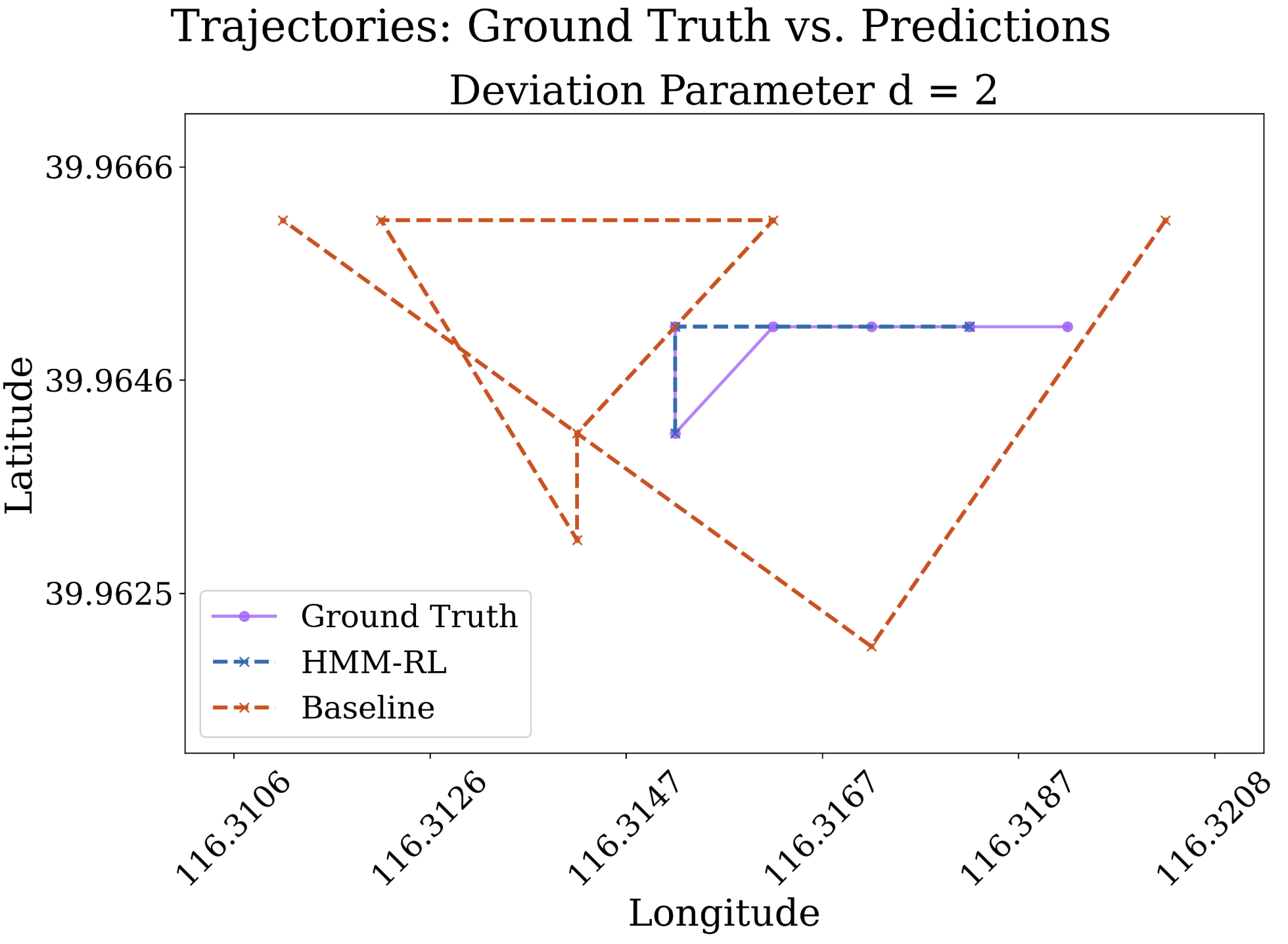}
    \caption{Geographic Visualization of One Trajectory in Geolife with Deviation Parameter $d=2$.}
    \label{fig:one traj}
  \end{minipage}\hfill
  \begin{minipage}[b]{0.32\textwidth}
    \centering
    \includegraphics[width=\linewidth]{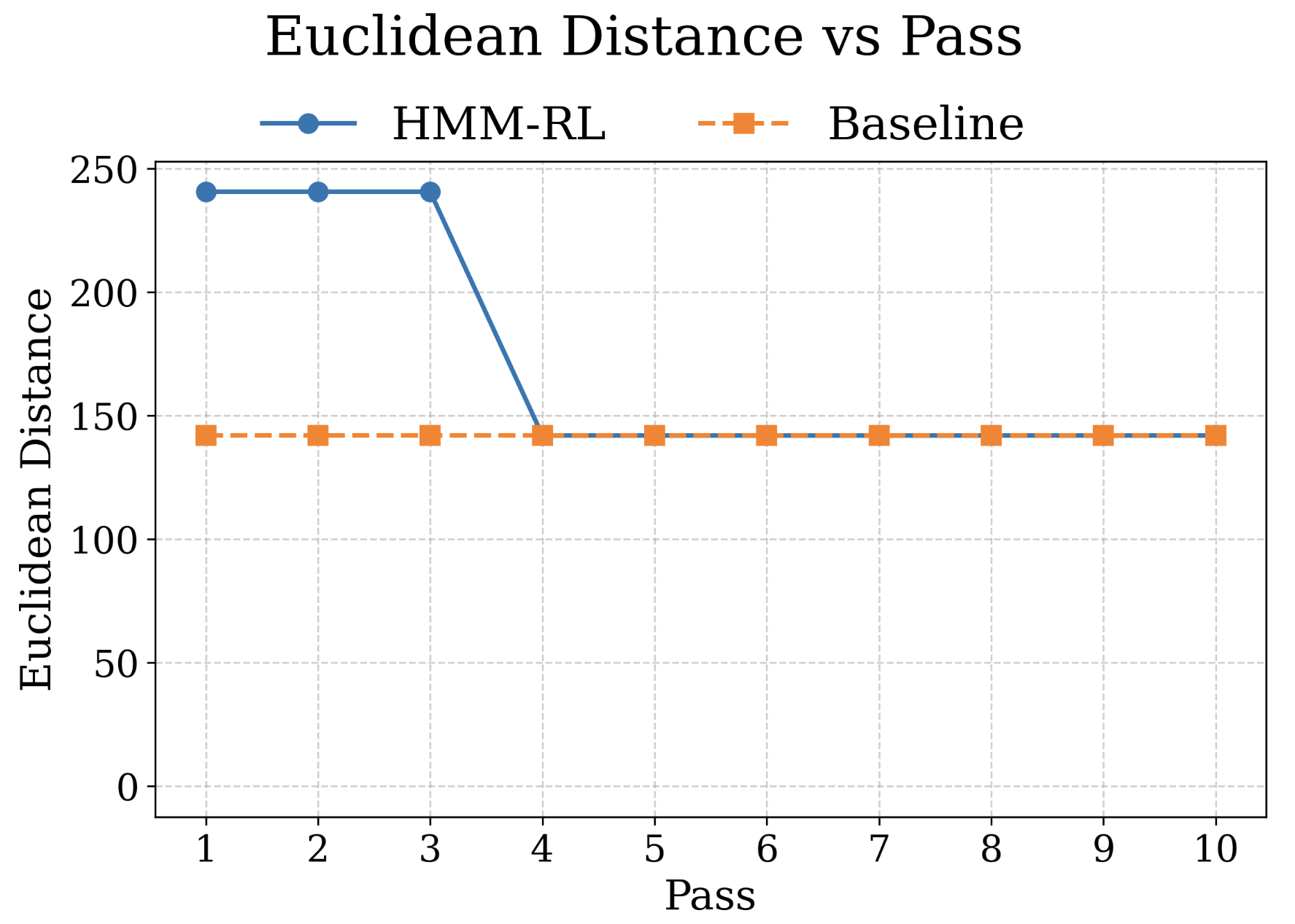}
    \caption{Euclidean distance between the predicted and ground-truth true locations of one trajectory without EPRL across 10 passes for the Geolife dataset.}
    \label{fig:without eprl}
  \end{minipage}\hfill
  \begin{minipage}[b]{0.32\textwidth}
    \centering
    \includegraphics[width=\linewidth]{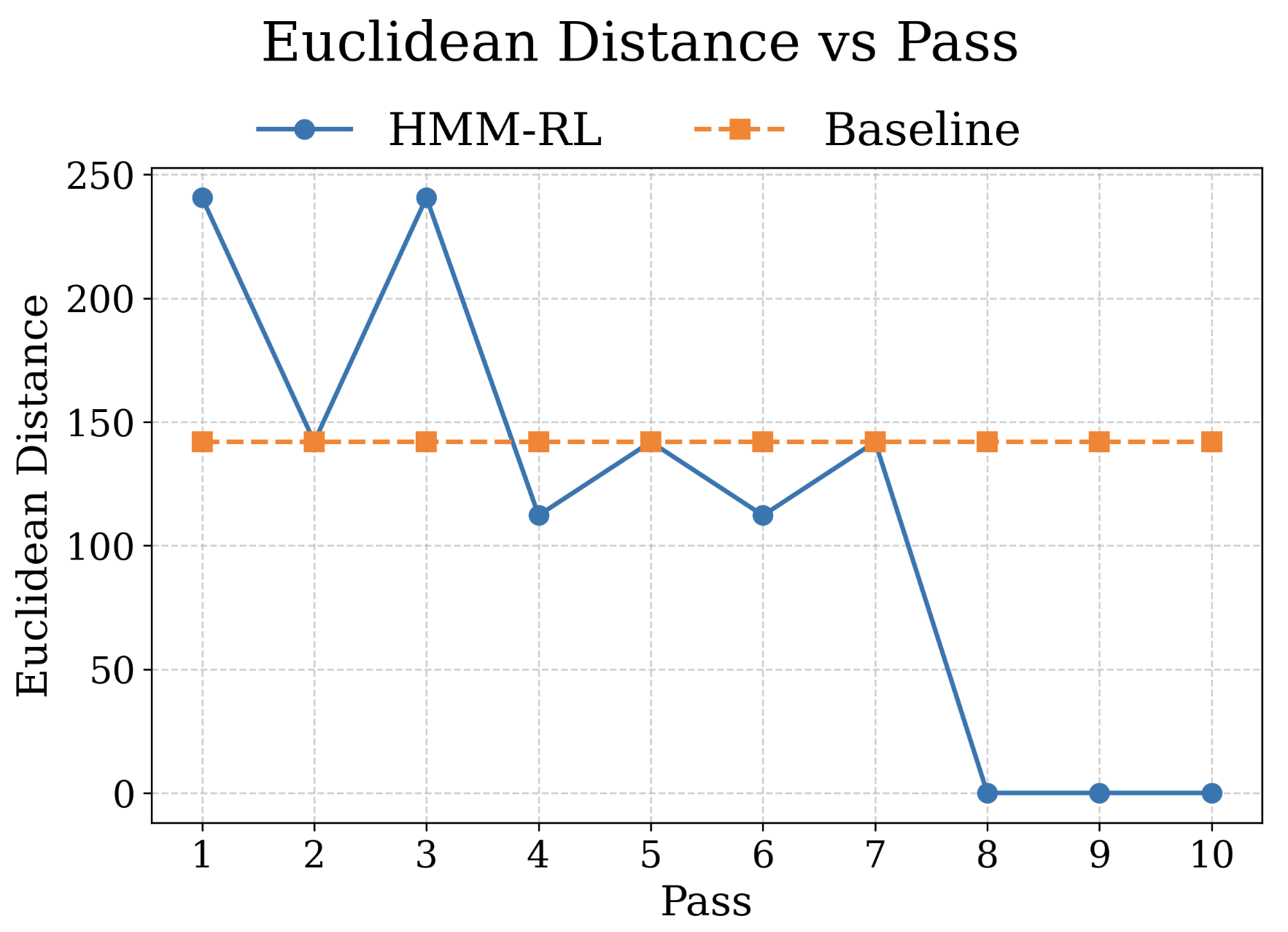}
    \caption{Euclidean distance between the predicted and ground-truth true locations of one trajectory with EPRL across 10 passes for the Geolife dataset.}
    \label{fig:with eprl}
  \end{minipage}
  \label{fig:combined}
\end{figure*}

\subsubsection{Effectiveness of EPRL}
We also compare HMM-RL with and without reinforcing the emission probability matrix when $R_{i-1} < \delta$ (EPRL), as elaborated in Section~\ref{subsec:rl based approach}. As indicated in Table~\ref{table:overall results}, the framework with EPRL consistently outperforms the one without across all datasets and metrics. Specifically, EPRL reduces A$^2$ED and AMED by 32.77\% and 23.06\%, respectively, demonstrating its effectiveness.

% baseline of 40.267 meters in A$^2$ED and 107.318 meters in AMED compared to the baseline, and 110.306 meters in A$^2$ED and 133.335 meters in AMED relative to the EPRL-enhanced variant. 
To further investigate the impact of EPRL, we analyze the evolution of Euclidean distance error over multiple passes on the Geolife dataset (Figures~\ref{fig:without eprl} and ~\ref{fig:with eprl}). As indicated in Figure~\ref{fig:without eprl}, we can see that, without EPRL, the error remains nearly constant across passes, indicating that the model fails to learn effectively when $R_{i-1} < \delta$, as no meaningful reinforcement updates occur, confirming our claims in Section~\ref{subsec:rl based approach}. In contrast, with EPRL enabled as indicated in Figure~\ref{fig:with eprl}, the error decreases across passes, confirming that emission probability reinforcement provides valuable feedback and promotes better convergence.

\subsection{Sensitivity Analysis}
\label{sec: sensitivity analysis}

We now examine the impact of key hyperparameters on attack performance, including the deviation parameter $d$, the privacy threshold $\lambda$, and attack model's hyperparameters $\gamma$ (restricting possible published region size), $k$ (controlling the size of the sliding window) and $\delta$ (controlling the threshold to reward in reinforcement learning). This analysis provides insight into how individuals can strengthen privacy guarantees when releasing data, and highlights trade-offs between privacy protection and data utility.

\subsubsection{Effect of Deviation Parameter $d$}

The deviation parameter $d$ controls the relative position of the true location within the published region. When $d=0$, the true location is centered in the published region; as $d$ increases, the true location moves farther from the center.  

We evaluate the effect of $d$ on attack performance. For $\lambda=0.1$, results are meaningful only up to $d=2$, since larger deviations may place the true location outside the published region. For example, with a published region of size $3 \times 5$, shifting by $3$ grids in either direction would move the true location outside the region. Experimentally, we find that 28.5\% of simulated true locations remain valid under $d=3$, compared with 100\% validity for $d=0,1,2$.

\begin{table*}[t!]
\centering
\caption{Effectiveness of Bi-Directional HMM-RL Algorithm in trajectory attacks, measured using (a) aggregate maximum Euclidean distance (AMED) error and (b) aggregate average Euclidean distance (A$^2$ED) error. \textbf{Bolded} entries denote the smaller error between the baseline and HMM-RL algorithm.}
\label{table:traj results}
% \renewcommand{\arraystretch}{0.95}
% \small
% \setlength{\tabcolsep}{4pt}
\begin{subtable}[t]{0.6\textwidth}
%\small
\centering
\caption{AMED error}
\label{table:traj-results-max}
\begin{tabular}{ccccccc} 
\toprule
& \multicolumn{3}{c}{Empirical Error} & \multicolumn{3}{c}{Theoretical Error} \\
\cmidrule(lr){2-4} \cmidrule(lr){5-7}
Model & $d=0$ & $d=1$ & $d=2$ & $d=0$ & $d=1$ & $d=2$ \\
\midrule
\multicolumn{7}{l}{\textbf{Syn-Chengdu}} \\
Baseline & 345.498 & \textbf{306.495} & 396.827 & 584.202 & 681.569 & 778.936 \\
\rowcolor{gray!20}
HMM-RL & \textbf{275.158} & 351.691 & \textbf{388.490} & 584.202 & 681.569 & 778.936 \\
\midrule
\multicolumn{7}{l}{\textbf{Syn-Xi’an}} \\
Baseline & 342.390 & \textbf{296.769} & \textbf{383.628} & 584.874 & 682.353 & 779.832 \\
\rowcolor{gray!20}
HMM-RL & \textbf{243.247} & 301.984 & 389.179 & 584.874 & 682.353 & 779.832 \\
\midrule
\multicolumn{7}{l}{\textbf{Geolife}} \\
Baseline & 431.311 & 438.122 & 532.337 & 596.298 & 695.681 & 795.064 \\
\rowcolor{gray!20}
HMM-RL & \textbf{268.210} & \textbf{320.830} & \textbf{427.527} & 596.298 & 695.681 & 795.064\\
\midrule
\multicolumn{7}{l}{\textbf{Porto Taxi}} \\
Baseline & \textbf{397.035} & 566.226 & \textbf{745.935} & 893.742 & 1042.699 & 1191.656 \\
\rowcolor{gray!20}
HMM-RL & 464.825 & \textbf{560.373} & 749.464 & 893.742 & 1042.699 & 1191.656 \\
\bottomrule
\end{tabular}
\end{subtable}
\hfill
\begin{subtable}[t]{0.38\textwidth}
\centering
\caption{A$^2$ED error}
\label{table:traj-results-avg}
%\small
\begin{tabular}{cccc} 
\toprule
& \multicolumn{3}{c}{Empirical Error} \\
\cmidrule(lr){2-4}
Model & $d=0$ & $d=1$ & $d=2$ \\
\midrule
\multicolumn{4}{l}{\textbf{Syn-Chengdu}} \\
Baseline & 192.737 & 222.961 & 284.674 \\
\rowcolor{gray!20}
HMM-RL & \textbf{131.684} & \textbf{182.640} & \textbf{195.987} \\
\midrule
\multicolumn{4}{l}{\textbf{Syn-Xi’an}} \\
Baseline & 188.501 & 217.460 & 277.632 \\
\rowcolor{gray!20}
HMM-RL & \textbf{113.419} & \textbf{154.298} & \textbf{197.546} \\
\midrule
\multicolumn{4}{l}{\textbf{Geolife}} \\
Baseline & 188.101 & 208.331 & 264.563 \\
\rowcolor{gray!20}
HMM-RL & \textbf{112.761} & \textbf{145.743} & \textbf{204.068} \\
\midrule
\multicolumn{4}{l}{\textbf{Porto Taxi}} \\
Baseline & 305.216 & 362.715 & 461.148 \\
\rowcolor{gray!20}
HMM-RL & \textbf{213.122} & \textbf{264.917} & \textbf{360.259} \\
\bottomrule
\end{tabular}
\end{subtable}
\end{table*}

\paragraph{Theoretical Worst-Case Results}
We first theoretically analyze the worst-case scenario, which is the largest distance between each predicted true location and its associated ground-truth true location. In trajectory attacks, each true location consists of exactly one grid cell, so the minimum published region size is $\ell = 1/\lambda = 10$ when $\lambda = 0.1$. The worst-case scenario occurs when the published region has shape $1\times \ell$ or $\ell\times 1$ in the grid space, with the predicted true location being at one of the endpoints. When $d=0$, $TL_i$ is at the center of the published region, so the associated theoretical maximum predicted error is
% The worst-case scenario maximizing the distance between the center and any endpoint arises when the published region has shape $1 \times L$. Following the expansion strategy in Section~\ref{sec:experimental setup}, the penultimate region must have size at most $\ell-1$. Thus, the worst case occurs when a $1 \times (\ell-1)$ region is expanded to $1 \times (\ell+1)$, with $\widehat{TL}_i$ at the edge. Because the Bi-Directional HMM-RL algorithm enforces $\widehat{TL}_i \subseteq PR_i$ for all $i \in [1,T]$, the theoretical AMED when $d=0$ is  
\(
\Big\lceil \tfrac{\ell+1}{2} \Big\rceil \times g = (6 \times g) \text{ meters},
\) 
where $g$ denotes the side length of the grids in each dataset (for example, $g=99.383$ for the Geolife dataset).

For a general $d \geq 0$, the theoretical maximum prediction error is 
\(
\Big(\Big\lceil \tfrac{\ell+1}{2} \Big\rceil + d\Big) \times g = ((6+d)\times g) \text{ meters}.
\)
This indicates that larger $d$ increases the theoretical maximum prediction error and thus strengthens privacy protection.  

% In contrast, the baseline model provides no guarantee on $\widehat{TL}_i$, and its worst-case theoretical maximum prediction error equals the diagonal of the bounding box, (i.e., 12377 meters for Syn-Chengdu, 12148 meters for Syn-Xi'an, 6521 meters for Geolife, and 6437 meters for Porto Taxi).

\paragraph{Experimental Results} 

\begin{figure*}[t]
    \centering
    \begin{subfigure}[b]{0.48\textwidth}
        \centering
        \includegraphics[width=\textwidth]{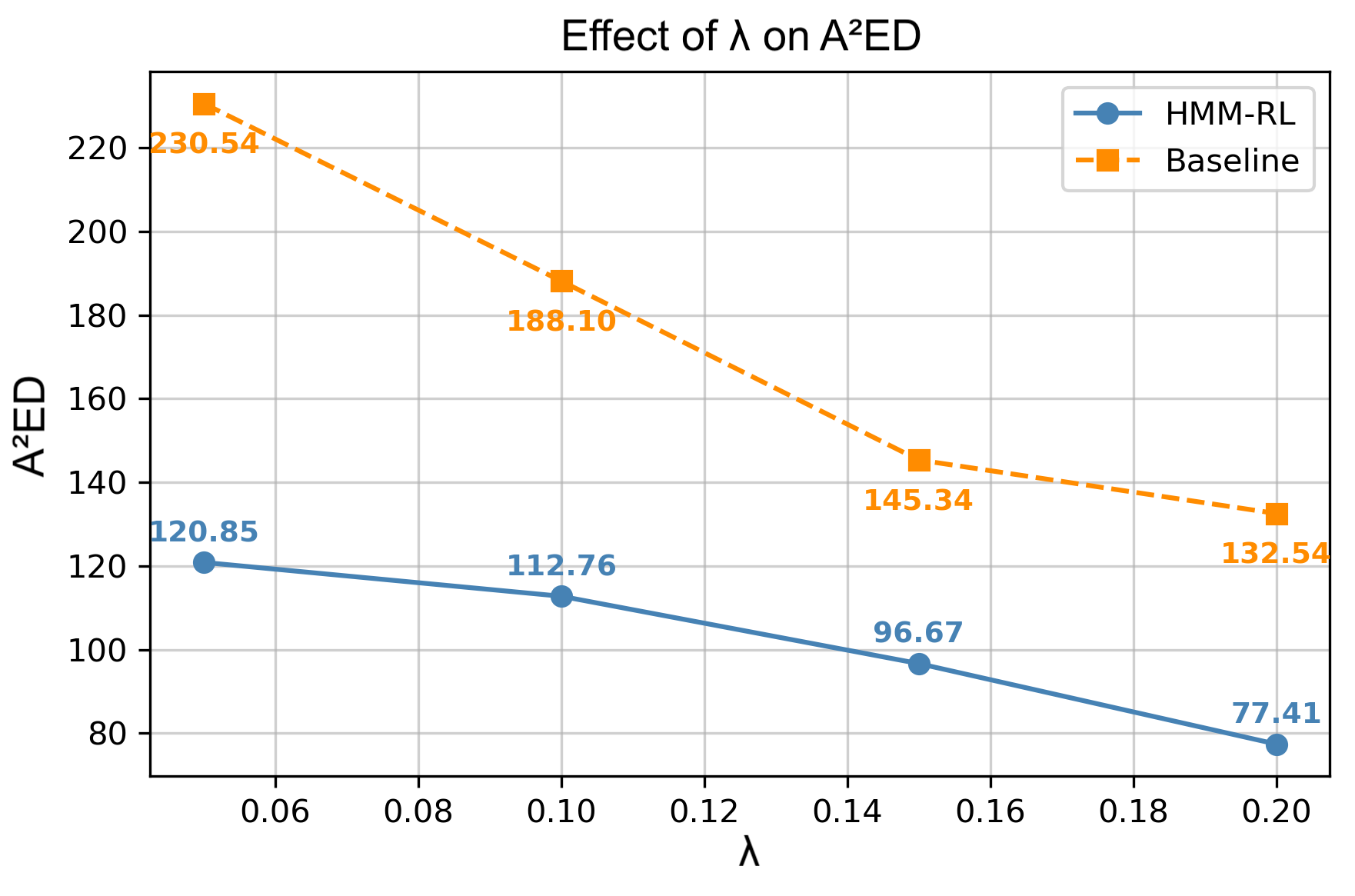}
        \caption{A$^2$ED}
        \label{fig:lambda AED sensitivity analysis}
    \end{subfigure}
    \hfill
    \begin{subfigure}[b]{0.48\textwidth}
        \centering
        \includegraphics[width=\textwidth]{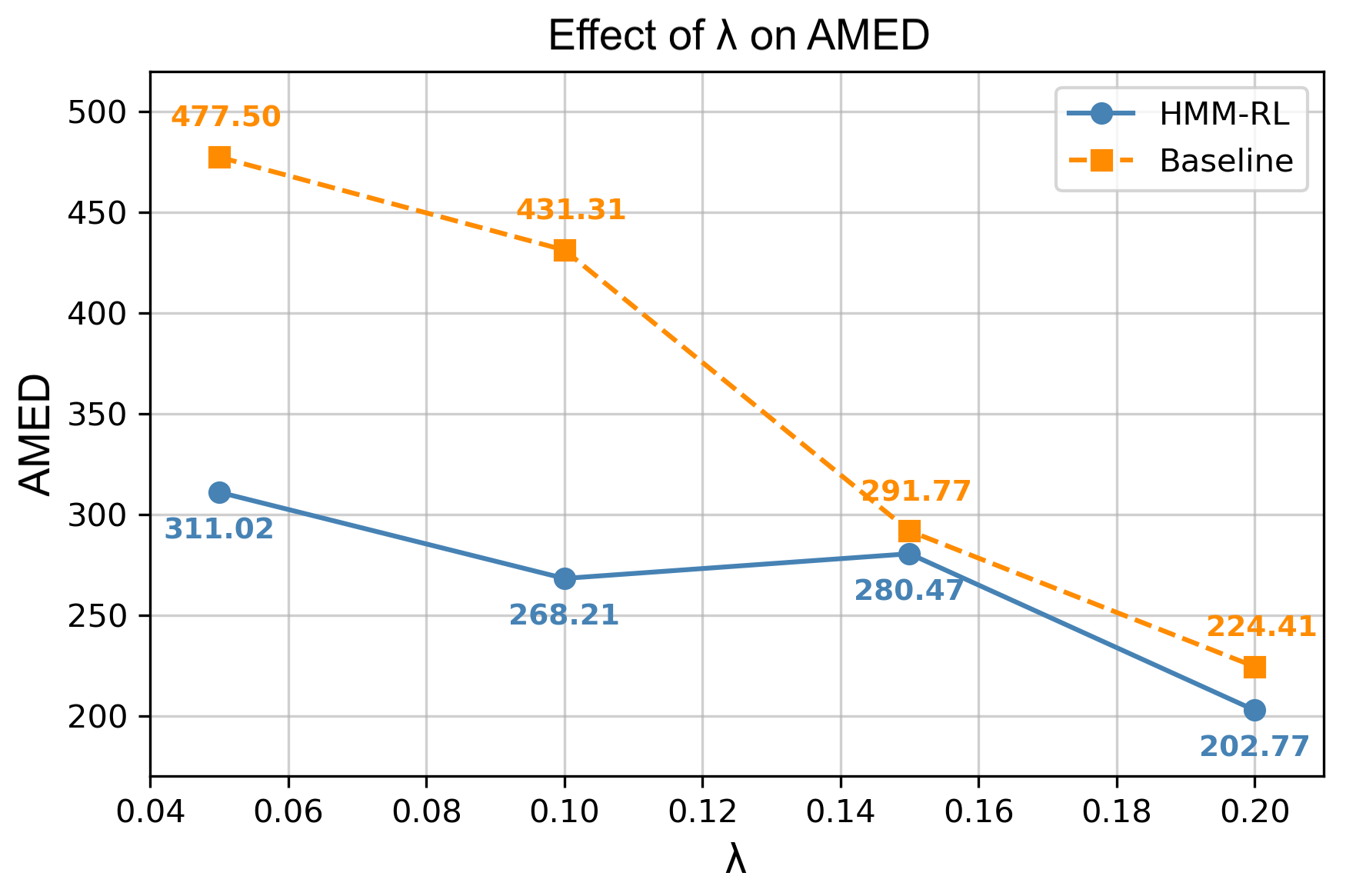}
        \caption{AMED}
        \label{fig:lambda AMED sensitivity analysis}
    \end{subfigure}
    \caption{Impact of the privacy threshold $\lambda$ on (a) aggregate average Euclidean distance (A$^2$ED) and (b) aggregate maximum Euclidean distance (AMED) for the Bi-Directional HMM-RL algorithm compared to the baseline.}
    \label{fig:lambda sensitivity analysis}
\end{figure*}

We generate published regions using $d \in \{0, 1, 2\}$. As shown in Table~\ref{table:traj results}, the Bi-Directional HMM-RL algorithm consistently achieves lower A$^2$ED across all settings, outperforming the baseline in 35.43\%, 26.03\%, and 26.19\% on average when $d = 0, 1, 2$, respectively. The same trend holds for AMED, where the Bi-Directional HMM-RL algorithm outperforms the baseline in 17.51\%, 2.83\%, and 4.97\% on average when $d=0,1,2$. Although in a few instances the baseline slightly outperforms the HMM-RL algorithm, its improvement is marginal, only 25.46 meters on average across 5 instances, compared with the larger gains achieved by the proposed method in the remaining cases with 81.27 meters on average across 7 instances.

Second, we find that our model's prediction errors increase as $d$ grows. For example, in the Geolife dataset, the AMED rises by 159.317 meters (approximately 1.6 grid cells) from $d = 0$ to $d = 2$, while the A$^2$ED increases by 91.307 meters. Recall that the attacker assumes $d = 0$ in the attack model when generating $\widehat{PR}_i$. Consequently, the attacker performs best when their assumed T2P mapping aligns with that of the data publisher, but less effectively when the mappings diverge, i.e., when $d$ increases from 0 to larger values. This suggests an interesting future direction where privacy can be enhanced by adopting T2P mappings that are less predictable to potential adversaries. However, even with the growth of error, our method still outperforms the baseline, like when d = 2, our method gives a decrease of 26.02 meters in AMED error and 82.54 meters in A$^2$ED error.

In short, despite the relative increase in prediction errors as $d$ grows, the results in Table~\ref{table:traj results} still demonstrate that our approach remains effective when the publisher's T2P mapping is not strictly deterministic. In practice, publishers often employ randomized or heuristic strategies when generating published regions. Our experiments model this by using deterministic expansion around the true location combined with a small random shift ($d$) to introduce variability. The results show that the HMM-based model can still learn accurate transition and emission patterns under such conditions, indicating that the attacker's inference capability is robust even against realistic, non-deterministic publishing mechanisms.

\subsubsection{Effect of Privacy Threshold $\lambda$}

The privacy threshold $\lambda$ constrains the attacker’s maximum confidence in identifying the true location. Smaller $\lambda$ enforces stronger privacy but reduces utility, as it requires a larger published region to hide the true location. We vary $\lambda$ from $0.05$ (lower bound size $20$) to $0.20$ (lower bound size $5$), with results on the Geolife dataset shown in Figure~\ref{fig:lambda sensitivity analysis}.  

Both models show that lower $\lambda$ values yield larger A$^2$ED and AMED, as expected. For example, at $\lambda=0.05$, at least $20$ grids must be published, which substantially reduces utility but provides stronger privacy. As $\lambda$ increases, the published regions shrink, improving utility at the cost of easier attacks. This illustrates the core privacy-utility tradeoff: reducing $\lambda$ strengthens privacy but degrades downstream usability.

\subsubsection{Effect of Attack Model’s Hyperparameters $\gamma, k, \delta$}

\begin{figure*}[t]
    \centering
    \includegraphics[width=\textwidth]{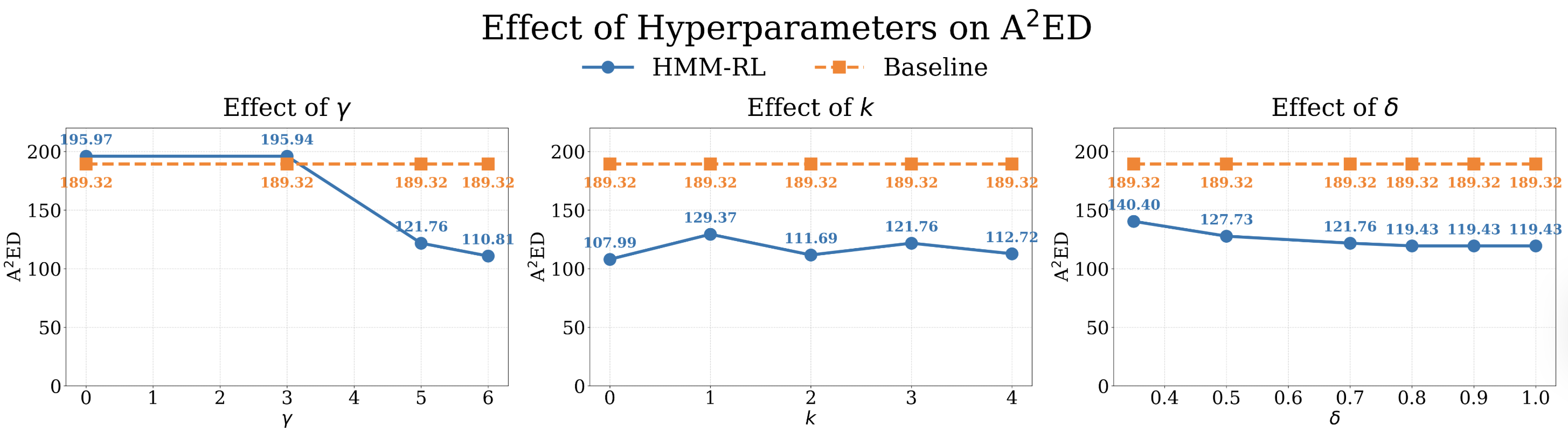}
    \caption{Aggregate average Euclidean distance (A$^2$ED) of Bi-Directional HMM-RL and baseline methods across attacker parameters ($\gamma, k, \delta$) on the Geolife dataset.}
    \label{fig:dbp results hyperparameters}
\end{figure*}

The attack model includes three hyperparameters: $\gamma$ (upper bound on published region size), $k$ (sliding-window range), and $\delta$ (reinforcement learning update threshold). Figure~\ref{fig:dbp results hyperparameters} shows the impact of these hyperparameters on our proposed model performance on Geolife, with the baseline as a reference.

The best A$^2$ED values are obtained when $\gamma$ is larger, indicating that the attack model performs best when it considers a broader range of possible published regions. However, practical computational and utility constraints limit the feasibility of including excessively large regions. 

The sliding-window size $k$ has relatively little influence, with A$^2$ED values varying only between 107.99 and 129.37.

The threshold parameter $\delta$ exhibits a monotonic relationship with A$^2$ED: as $\delta$ increases, A$^2$ED decreases, reflecting improved learning performance. Beyond $\delta > 0.8$, performance converge, with A$^2$ED stabilizing around 119.43, suggesting diminishing returns from further tightening the threshold. This trend aligns with the learning dynamics of the HMM-RL model, where $\delta$ governs the evaluation of predicted published regions at each time step. When $\delta$ is too relaxed (e.g., $\delta = 0.3$), reinforcement learning provides limited benefit, as most predictions are accepted regardless of accuracy. In contrast, stricter thresholds (e.g., $\delta > 0.7$) ensure that reinforcement updates are triggered only by sufficiently accurate predictions, allowing the model to extract more meaningful feedback and achieve better convergence.

\section{Conclusion and Future Work}
\label{section: conclusion}

In this work, we introduced a novel attack model that exposes privacy vulnerabilities in sequential data releases, even when each individual release independently satisfies privacy constraints. By exploiting temporal dependencies through a bi-directional Hidden Markov Model enhanced with reinforcement learning, our approach enables adversaries to infer sensitive information -- such as individual trajectories -- with significantly higher accuracy. Experiments on both real-world and synthetic datasets demonstrate that our model consistently outperforms existing baselines that treat sequential releases as independent. These findings highlight an important and underexplored threat vector in privacy-preserving data publishing and open several promising avenues for future research.  

\textbf{Sequential-dependency-aware privacy mechanisms.}  
Our results suggest that traditional privacy guarantees must be reconsidered in settings involving temporally correlated data releases. Future work should explore defense strategies that explicitly account for sequential dependencies. Potential directions include modifying the publication process by lowering the privacy threshold $\lambda$ to enlarge anonymized regions, designing trajectory-to-publication (T2P) mappings that intentionally deviate from attacker assumptions, or adopting uniform publication strategies that minimize distinguishability among trajectories. Another promising line of research is to generalize our attack framework to predictive or adaptive adversaries with varying levels of background knowledge.  

\textbf{Enhancing existing privacy-preserving frameworks.}  
Well-established methods such as $k$-anonymity and differential privacy could be extended to incorporate temporal correlations. Developing a \emph{temporal differential privacy} framework or other sequence-aware protection schemes could offer stronger resistance to cross-release inference. Adaptive mechanisms that dynamically adjust privacy threshold $\lambda$ -- for instance, increasing injected noise or expanding published regions when correlations are strong -- represent another promising approach. Such mechanisms could leverage online learning or reinforcement learning to estimate real-time inference risks and automatically tune privacy parameters. Furthermore, publishers could obscure temporal information by releasing time ranges rather than exact timestamps to further mitigate linkage attacks.  

\textbf{Beyond trajectory data.}  
Although our study focuses on mobility trajectories, the identified risks extend to other domains that involve temporally correlated data. In healthcare, for example, sequential releases of patient information may reveal disease progression or treatment responses; in finance, repeated transaction disclosures could expose behavioral patterns over time. Applying and benchmarking our attack model in such domains would provide a more comprehensive understanding of privacy degradation under temporal dependence and guide the design of domain-specific defense mechanisms.  

Overall, this work underscores the importance of rethinking privacy guarantees in dynamic, sequential settings and provides a foundation for developing more resilient privacy-preserving mechanisms in the era of continuous data generation and release.

\section*{AI-Generated Content Acknowledgement}
Language models were used to check for grammatical mistakes. Language models were also used to select appropriate wordings and improve sentence flow.

\bibliographystyle{IEEEtran}   % numeric IEEE bibliography style
\bibliography{IEEEabrv}
\end{sloppy}
\end{document}